\documentclass[a4paper]{article}
\pdfoutput=1

\usepackage{graphicx}
\usepackage[]{natbib}
\usepackage{verbatim}
\usepackage{amsmath}
\usepackage{amsthm}
\usepackage{amsfonts}
\usepackage{authblk}
\usepackage{array}
\newcolumntype{R}[1]{>{\raggedleft\let\newline\\\arraybackslash\hspace{0pt}}p{#1}}
\usepackage[official]{eurosym}
\usepackage[english]{babel}
\usepackage[utf8]{inputenc}
\usepackage[ruled,linesnumbered]{algorithm2e}
\usepackage{tikz}
\usetikzlibrary{shapes,arrows, shapes.geometric, shapes.symbols}

\theoremstyle{definition}
\newtheorem{definition}{Definition}

\usepackage{tikz}
\usetikzlibrary{shapes,arrows, shapes.geometric, shapes.symbols}
\tikzstyle{every label}= [black]
\tikzstyle{place}=[circle,draw=black,minimum size=6mm, node distance=2cm]
\tikzstyle{transition}=[rectangle,draw=black, minimum size=6mm, node distance=2cm]
\tikzstyle{pre}=[<-,shorten <=1pt,>=stealth',semithick]
\tikzstyle{post}=[->,shorten >=1pt,>=stealth',semithick]
\begin{document}
\bibliographystyle{model1-num-names}

\title{Fitting random cash management models to data}

\author[1]{Francisco Salas-Molina\footnote{Corresponding author. E-mail addresses: \textit{frasamo@upv.es}}}
	
\affil[1]{Universitat Polit\`enica de Val\`encia, Ferrandiz y Carbonell s/n, 03801 Alcoy, Spain}

\maketitle

\begin{abstract}
Organizations use cash management models to control balances to both avoid overdrafts and obtain a profit from short-term investments. Most management models are based on control bounds which are derived from the assumption of a particular cash flow probability distribution. In this paper, we relax this strong assumption to fit cash management models to data by means of stochastic and linear programming. We also introduce ensembles of random cash management models which are built by randomly selecting a subsequence of the original cash flow data set. We illustrate our approach by means of a real case study showing that a small random sample of data is enough to fit sufficiently good bound-based models.
\\
\\

\\
\end{abstract}

\section{Introduction\label{intro}}

A wide range of economic organizations manage cash for operational, precautionary and speculative purposes \citep{keynes1936general}. Cash management models help decision-makers in their daily job of controlling cash balances. Since the seminal works by \cite{baumol1952transactions} and \cite{miller1966model}, cash management models follow a inventory control approach. Within this framework, cash balances are allowed to wander around until some control bounds, usually a higher bound and a lower bound, are reached. Then, a control action is made to restore the balance to a given target level. The set of control actions deployed over a period of time is called a policy and it is usually determined by simple rules derived from the set of bounds of the model. 

The bound-based control approach is based on the strong assumption of a particular probability distribution for cash flows, which is usually assumed to be a normal, independent and stationary cash flow as in \cite{miller1966model,baccarin2009optimal,premachandra2004diffusion}. Surprisingly, the use of empirical data sets in cash management research is limited to recent contributions such as \cite{gormley2007utility} and \cite{salas2017empowering}, in which alternative forecasters are used to obtain predictions as a key input to cash management models, and \cite{salas2016multi}, in which a multiobjective approach to the cash management problem is proposed. We here follow a different data-driven approach. Instead of fitting forecasters as in \cite{gormley2007utility} and \cite{salas2017empowering}, we here fit cash management models. Furthermore, \cite{gormley2007utility} proposed a bound-based model using forecasts as a key input and they used genetic algorithms to find sufficiently good bounds. On the other hand, \cite{salas2016multi} used simulation techniques and the \cite{miller1966model} model within a multiobjective framework considering not only the cost but also the risk of alternative policies. In this paper, we first propose a mixed-integer linear program that derives from a general stochastic programming problem to obtain a bound-based model from available data. In order to avoid the computational burden of possibly very large mixed-integer linear programs, we propose a method based on random cash flow subsequences.

We here follow a data-driven approach that mainly focuses on models for cash managers. However, we rely on several machine learning concepts to build our approach. Machine learning covers a wide range of algorithms that can learn from data to make better decisions. A paradigmatic case is deep learning because it requires very little engineering and it can take advantage of computational power and data availability \citep{lecun2015deep,schmidhuber2015deep}. Deep learning models are built through multiple processing layers that are able to discover how computers should modify or adapt their actions to be more precise. On the other hand, random forests are also an interesting technique based on an ensemble of slightly different decision trees \citep{ho1998random,breiman2001random}. The output of the global model is then obtained by combining the output of multiple randomly trained models. In this paper, we use this particular feature to fit random cash management models.

Fitting a model to data means finding the best parameters of the model according to an objective function and a given data set. We here describe a general procedure to fit a set of control bounds to a given data set of cash flows. We make no assumption on the form of the cash flow process under consideration. Then, our approach accepts as a key input a data set with either: (i) past cash flow observations; (ii) cash flows sampled from a particular probability distribution; or (iii) a set of cash flow predictions. Similarly to the ordinary least squares method used in linear regression, we rely on an optimization procedure to produce the solution that minimizes some objective function for a given data set. Instead of a data set with previous examples, we use a data set with cash flow observations. Instead of minimizing the sum of squared deviations, we minimize the sum of costs. And instead of obtaining a set of regression coefficients, we obtain a set of control bounds ready to be used by cash managers. 

Relevant related works are those based on stochastic programming (SP) to address different cash management problems as in \cite{golub1995stochastic,gardin1995liquidity,gondzio2001high} and \cite{castro2009stochastic}. In these works, the authors rely on SP and a set of previous realizations of random variables to improve current techniques to manage cash (e.g. in automatic teller machines in the case of \cite{castro2009stochastic}). Thus, this paper is an extension of this body of SP works to fit bound-based models in cash management. A further advantage of our proposal is that the solution provided is a cash management model of the \cite{miller1966model} type. In other words, we provide cash managers with a set of simple decision rules based on a set of control bounds that proved to be optimal for a given cash flow data set. This fact avoids the drawback of solving a new (possibly large) problem at each time step when information about a new initial condition is available according to the so-called receding horizon philosophy \citep{bemporad1999control,camacho2007model}. In addition, our approach can be extended to fit other cash management models such as the one proposed by \cite{stone1972use} or by \cite{gormley2007utility}.

In order to speed up computations, we describe a procedure to randomly select a subsequence of the original cash flow data set to construct an ensemble of random cash management models. The policy to deploy is elicited by averaging the output of randomly trained models similarly to the methods used in machine learning to train random forests \citep{ho1998random,breiman2001random}. To illustrate our approach, we present a case study with real data from an industrial company in Spain. As a benchmark, we use the equations proposed by the \cite{miller1966model} model to obtain a set of three bounds. The reason to select this model is twofold. First, its relevance. The Miller and Orr model was the first stochastic cash management model and it has become a framework for subsequent research in cash management (some recent examples are \cite{premachandra2004diffusion,da2014evolutionary}). Second, its simplicity. While other models propose a higher number of bounds (see e.g. \cite{eppen1969cash,stone1972use}), the Miller and Orr model is based on only three control bounds allowing us to limit the number of decision variables and constraints for illustrative purposes.

Summarizing, we propose a general methodology to fit cash management models based on control bounds to data as a feasible way to solve the cash management problem (CMP). More precisely, we highlight three main contributions:
\begin{enumerate}
\item We provide a method to solve the CMP when using bound-based models without making any assumption on the underlying cash flow process.
\item We construct data-driven cash management models by fitting parameters to data.
\item We introduce ensembles of random cash management models.
\end{enumerate}

This paper is organized as follows. In Section \ref{sec:background}, we provide useful background on the Miller and Orr model, the usual cost functions used in cash management and about stochastic programming in cash management. In Section~\ref{sec:datadriven}, we introduce our method to fit models to cash flow data sets. Next, we present a case study using real data in Section~\ref{sec:case}. Finally, we conclude in Section \ref{sec:conclusions} suggesting natural extensions of our work.

\section{Background\label{sec:background}}

In this section, we first provide useful background on the Miller and Orr model that we later use in a case study to illustrate our data-driven approach. Next, we describe the usual cost functions used in cash management. Finally, we describe a related approach to cash management based on stochastic programming.

\subsection{The Miller and Orr model}

\begin{figure}[htb]
\centering
\begin{tikzpicture}[bend angle=45,node distance = 1.5cm]
\node[place] (account)  {$1$};
\node [text centered, above of=account] (f1) {$f_t$}
	edge[post] (account);  
\node[place] (invest_account) [right of= account]  {$2$}
	edge[post] node [above] {$x_{t}$} (account);
\end{tikzpicture}
\caption{\label{fig:system}A cash management system with two assets.}
\end{figure}
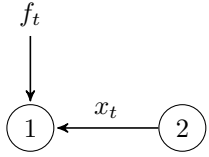

Consider a cash management system for a typical company as shown in Figure~\ref{fig:system}. This system comprises two accounts (depicted as circles), a control action $x_t$ between accounts and an external cash flow $f_t$ summarizing both inflows from debtors and outflows to creditors at each time step $t$. Cash managers can adjust cash balances $b_t$ in account~1 for operational purposes by selling available investments in account~2 through control action $x_t$ at a fixed cost $\gamma_0$. On the other hand, idle cash balances in account~1 can be allocated in investment account~2 in exchange for a given return $v$ per money unit when $x_t<0$. \cite{miller1966model} proposed a model to control balances for the two-assets system in Figure~\ref{fig:system} by assuming that stochastic cash flows $f_t$ are generated by a stationary random walk with standard deviation $\sigma$.

Under this framework, cash managers seek to find sequence $x=\{x_1, x_2, \ldots, x_\tau\}$ to minimize long-rung average daily cost of managing their cash balances over any planning horizon of $\tau$ days given by:
\begin{equation}
\operatorname{min} \left[ \gamma_0 \cdot \frac{\operatorname{E}(n)}{\tau} + v \cdot \operatorname{E}(b)\right]
\label{eq:obj1}
\end{equation}
where $\operatorname{E}(n)$ is the expected number of transfers $n=\{|x|:x_i \neq 0\}$ during planning horizon $\tau$, and $\operatorname{E}(b)$ is the average daily balance derived from sequence $b=\{b_1, b_2, \ldots, b_\tau \}$ containing cash balances at each time step. Following the recommendations in \cite{gormley2007utility}, we use an indicator function $I_q$ that takes value one when condition $q$ holds, zero otherwise, to rewrite objective function in equation \eqref{eq:obj1} as follows:
\begin{equation}
\operatorname{min} \left[ \frac{\gamma_0}{\tau} \sum_{t=1}^\tau{I_{x_t \neq 0}}  +\frac{v}{\tau} \sum_{t=1}^\tau {b_t} \right]
\label{eq:obj2}
\end{equation}
subject to the following state transition law with an initial state $b_0$:
\begin{equation}
b_t = b_{t-1} + f_t + x_t.   
\label{eq:cont}
\end{equation}

\cite{miller1966model} proposed a bound-based model by showing that the optimal policy (when cash flows follow a stationary random walk with standard deviation $\sigma$) is obtained by defining three control bounds $L$, $Z$ and $H$. These bounds allow determining specific control action $x_t$ and cash balance $b_t$ at each time step $t$ from a previous balance $b_{t-1}$ and an external cash flow $f_t$. Formally, control action $x_t$ is elicited by comparing the current cash balance derived from previous balance $b_{t-1}$ and actual cash flow $f_t$ to the lower and upper bounds as follows:
\begin{equation}
x_t =\left\{\begin{array}{lll}Z-b_{t-1}-f_t, & \mbox{if} & b_{t-1}+f_t  > H \\ 0, &\mbox{if} & L<b_{t-1}+f_t < H  \\Z-b_{t-1}-f_t, & \mbox{if} & b_{t-1}+f_t < L. \end{array}\right.
\label{eq:TransferMiller}
\end{equation}

\begin{figure}[!htb]
\centering
\includegraphics[scale=0.5]{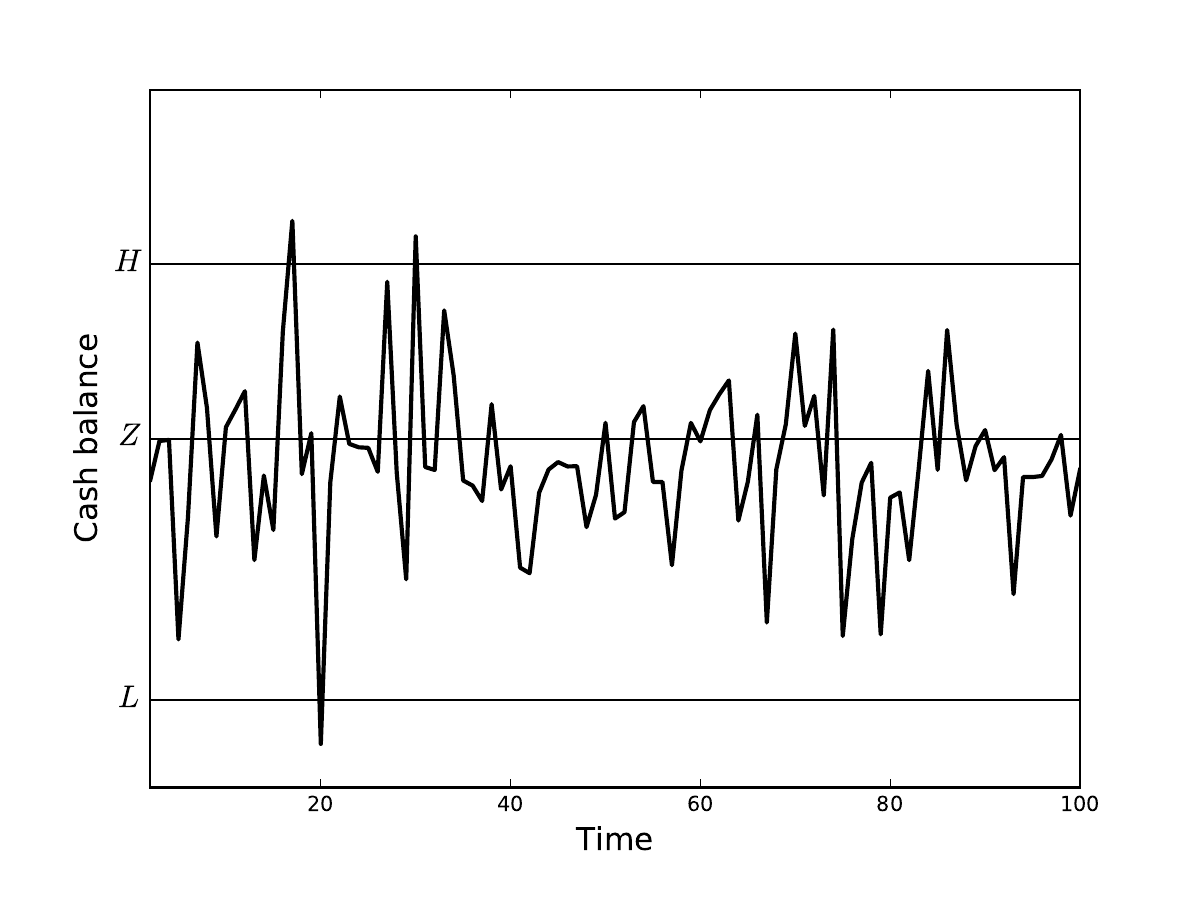}
\caption{\label{fig:MillerOrrModel}The Miller and Orr model.} 
\end{figure}

Although Miller and Orr set lower limit $L$ to zero in their work, a real cash manager should set a lower limit above zero for precautionary motives as recommended in \cite{ross2002fundamentals}. This lower limit represents a safety cash buffer and its selection will depend on the level of risk the company is willing to accept. As a result, when the cash balance reaches $L$, a positive transfer is made to restore the balance to $Z$ (from account~2 to account~1 in Figure \ref{fig:system}). Similarly, when $H$ is reached a negative transfer (from account~1 to account~2 in Figure \ref{fig:system}) is made to restore the balance to a target level $Z$ as shown in Figure~\ref{fig:MillerOrrModel}. After setting lower limit $L$ for precautionary purposes and by minimizing objective function \eqref{eq:obj2} with respect to policy $x$, \cite{miller1966model} showed that the optimal policy is given by equation \eqref{eq:TransferMiller} with parameters $Z$ and $H$ set as follows:
\begin{equation}
Z=L+\left(\frac{3 \cdot \gamma_0 \cdot  \sigma^2}{4 \cdot v}\right)^{1/3}
\label{eq:Z}
\end{equation}
and
\begin{equation}
H=3 \cdot Z-2 \cdot L.
\label{eq:H}
\end{equation}

The reasoning behind the optimality of these bounds requires rewriting objective function \eqref{eq:obj1} in terms of bounds $H$ and $Z$. By setting $R=H-Z$ and $L=0$, the problem can be stated in terms of the variance of the net cash flows as:
\begin{equation}
\min_{R,Z} \left\lbrace \frac{\gamma_0 \sigma^2}{RZ}+\frac{v(R+2Z)}{3} \right\rbrace.
\label{eq:obj4}
\end{equation}

The first term of objective function \eqref{eq:obj4} relates transaction cost $\gamma_0$ with the inverse value of the expected duration of a random walk of variance $\sigma$ starting at level $Z$ and ending at bounds $H$ or $L$. The second term relates the holding cost with the average cash balance given by $(H+Z)/3$. The necessary conditions for a minimum are that the partial derivatives of objective function \eqref{eq:obj4} with respect to $R$ and $Z$ are equal to zero that ultimately lead to the bound expressions described in equations \eqref{eq:Z} and \eqref{eq:H}. These results imply that the greater the transfer cost ($\gamma_0$), the higher the target cash balance ($Z$), and the greater the holding cost ($v$), the lower the target cash balance ($Z$). However, the greater the uncertainty of net daily cash flows, measured by $\sigma$, the higher the target cash balance ($Z$).

\subsection{Holding and transaction costs in cash management\label{sec:costs}}

In what follows, we consider a more general approach than \cite{miller1966model} with respect to cost functions as described in recent cash management works (see e.g. \cite{gormley2007utility,salas2016multi}). Any positive transaction adding cash to an account may have a cost, which may include a fixed part  ($\gamma_0^{+}$) and a variable part ($\gamma_{1}^{+}$). On the other hand, a negative transaction removing cash from an account may also have a cost with a fixed part  ($\gamma_0^{-}$) and a variable part ($\gamma_{1}^{-}$). Furthermore, at the end of the day, a holding cost~($v$) per money unit is charged if a positive cash balance occurs, or a penalty cost~($u$) per money unit is charged if a negative cash balance occurs. According to this cost structure, a general daily cost function is defined as:
\begin{equation}
c(x_t)=\Gamma(x_t)+\Lambda(b_t)
\label{eq:DailyCostFunction}
\end{equation}
where $\Gamma(x_t)$ is a transfer cost function, and $\Lambda(b_t)$ stands for a holding/shortage cost function. The transfer cost function $\Gamma(x_t)$ is defined as:
\begin{equation}
\Gamma(x_t)=\left\{\begin{array}{lll}\gamma_{0}^{-}-\gamma_{1}^{-}\cdot x_t & \mbox{if} & x_t<0, \\ 0 & \mbox{if} & x_t=0,  \\\gamma_{0}^{+}+\gamma_{1}^{+} \cdot x_t & \mbox{if} & x_t>0.\end{array}\right.
\label{eq:TransferFunction}
\end{equation}

Additionally, the holding/shortage cost function $\Lambda(b_t)$ is expressed as:
\begin{equation}
\Lambda(b_t)=\left\{\begin{array}{lll}-u\cdot b_t & \mbox{if} & b_t<0;u>0, \\v\cdot b_t & \mbox{if} & b_t \geq 0;v>0.\end{array}\right.
\label{eq:HoldingShortageCost}
\end{equation}

As a result, cash managers aiming to derive cash management policies need to solve the following program:
\begin{equation}
\operatorname{min} \frac{1}{\tau} \sum_{t=1}^\tau{\left[ (\gamma_{0}^{-} - \gamma_{1}^{-} \cdot  x_t) I_{x_t < 0} + (\gamma_{0}^{+} + \gamma_{1}^{+} \cdot x_t) I_{x_t > 0} + b_t (v \cdot I_{b_t \geq 0} - u \cdot I_{b_t < 0}) \right]}
\label{eq:obj3}
\end{equation}
subject to transition equation \eqref{eq:cont}. Note that the presence of indicator functions $I_q$ in objective function \eqref{eq:obj3} implies non-linearity since these functions depend on the value of decision variables $x_t$ and $b_t$. This fact complicates the selection of the best policies. In Section \ref{sec:datadriven}, we provide a method to derive cash management policies from data that relies on mixed integer linear programming to overcome this problem.

\subsection{Stochastic programing and cash management\label{sec:stochastic}}

Within a general formulation of a stochastic problem \citep{birge2011introduction}, we have to make decisions under some degree of uncertainty. These decisions are called first-stage decisions and are usually summarized in a vector $\boldsymbol{x}$ of decision variables. When new information on the realization
of some random vector $\boldsymbol{\xi}$ is available, second-stage decisions $\boldsymbol{y}$ are taken. The ultimate goal is to find decisions $\boldsymbol{x}$ that minimize average costs according to the distribution of $\boldsymbol{\xi}$ by means of the so-called stochastic program with recourse as follows:
\begin{equation}
\min_{\boldsymbol{x}} c(\boldsymbol{x}) + \operatorname{E}_{\boldsymbol{\xi}} \left[ Q(\boldsymbol{x},\boldsymbol{\xi}) \right]
\label{eq:sp}
\end{equation}
subject to:
\begin{equation}
A \boldsymbol{x} = \boldsymbol{d}
\end{equation}
\begin{equation}
\boldsymbol{x} \in \mathbb{R}^{n_1}, \hspace{2mm} \boldsymbol{x} \geq 0
\end{equation}
where:
\begin{equation}
Q(\boldsymbol{x},\boldsymbol{\xi}) = \min_{\boldsymbol{y}} \{q (\boldsymbol{y},\boldsymbol{\xi}) \hspace{2mm} |  \hspace{2mm} W\boldsymbol{y} = h(\boldsymbol{\xi}) - T(\boldsymbol{\xi}) \boldsymbol{x} \}
\end{equation}
\begin{equation}
\boldsymbol{y} \in \mathbb{R}^{n_2}
\label{eq:sp_end}
\end{equation}
where $q(\boldsymbol{y},\boldsymbol{\xi})$, $h(\boldsymbol{\xi})$ and  $T(\boldsymbol{\xi})$ are usually linear functions of random variable $\boldsymbol{\xi}$, matrix $W$ is assumed to be fixed, and $\operatorname{E}_{\boldsymbol{\xi}}$ denote mathematical expectation with respect to $\boldsymbol{\xi}$.

Cash management usually involves a sequence of decisions over a given planning horizon. Then, we need to formulate a multistage stochastic problem with $\tau$ time steps \citep{castro2009stochastic}:
\begin{equation}
\min_{\boldsymbol{x}_i} c_1(\boldsymbol{x}_1) + \operatorname{E}_{\xi^2} \left[ Q(\boldsymbol{x},\xi^{\tau}) \right]
\label{eq:sp2}
\end{equation}
subject to:
\begin{equation}
W_1 \boldsymbol{x}_1 = \boldsymbol{h}_1
\label{eq:c1sp2}
\end{equation}
\begin{equation}
W_2 \boldsymbol{x}_2(\xi^2) + T_1(\xi^2)\boldsymbol{x}_1 = \boldsymbol{h}_2(\xi^2)
\label{eq:c2sp2}
\end{equation}
\begin{equation}
W_i \boldsymbol{x}_i(\xi^i) + T_{i-1}(\xi^{i-1})\boldsymbol{x}_{i-1} = \boldsymbol{h}_i(\xi^i), \hspace{2mm} i = 3, 4, \ldots, \tau
\label{eq:c3sp2}
\end{equation}
\begin{equation}
\boldsymbol{x}_1 \geq 0,\hspace{2mm} \boldsymbol{x}_i(\xi_i) \geq 0, \hspace{2mm} i = 2, 3, \ldots, \tau,
\label{eq:sp2_end}
\end{equation}
where $\xi^i$ denotes the history of random events up to time step $i$, $W_i$ ($i=1, \ldots, \tau$) are known $m_i \times n_i$ matrices, $\boldsymbol{h}_1$ is a known vector in $\mathbb{R}^{m_1}$, $T_{i-1}$ are $m_i \times (n_i - 1)$ random matrices, $\boldsymbol{h}_i$ are random vectors in $\mathbb{R}^{m_1}$, and $\boldsymbol{x}_i$ ($i = 1, \ldots, \tau$) are vectors of decision variables at time step $i$ that depends on past events. In words, term $\operatorname{E}_{\xi^2} \left[ Q(\boldsymbol{x},\xi^{\tau}) \right]$ in objective function \eqref{eq:sp2} summarizes the expected cost from the second time step to the end of planning horizon $\tau$. In addition, constraints \eqref{eq:c1sp2}, \eqref{eq:c2sp2} and \eqref{eq:c3sp2} define the state transition between stages. In Section \ref{sec:datadriven}, we will see that our data-driven proposal is a special case of the general multistage stochastic problem encoded from equation \eqref{eq:sp2} to \eqref{eq:sp2_end}.

\section{A data-driven procedure to fit bound-based models\label{sec:datadriven}}

In this section, we introduce our data-driven stochastic approach to select the set of control bounds that determines the policy to be deployed by cash managers when using bound-based models. We first provide some useful definitions; next, we describe a novel method to fit bound-based cash management models; and finally, we introduce the concept of generalization power of cash management models.

\subsection{Some useful definitions\label{sec:definitions}}

As described in Section \ref{sec:costs}, cash managers make decisions under some economic context. 

\begin{definition}
A \textbf{cost structure} $\alpha$ is a tuple $\left(\gamma_0^+,\gamma_0^-,\gamma_1^+,\gamma_1^-,v,u\right)$ defining the transaction and holding costs for a given context.
\end{definition}

Within this context, we assume that cash managers have been able to observe the external net cash flows during a given period of time. Then, a data set of observed net cash flows is available as a key input to derive a cash management model.

\begin{definition}
Given a cash flow data set $\boldsymbol{f}=\{f_t:t=1,2,\ldots, N\}$ of size $N \in \mathbb{N}$, an initial condition~$b_0$, and a cost structure $\alpha$, a \textbf{cash management model} $h(\boldsymbol{f},b_0,\alpha)$ returns a policy that optimizes some objective function.
\end{definition}

By considering alternative cash flow data sets, cash managers can design a number of different cash management models to define a combined policy.

\begin{definition}
An \textbf{ensemble of cash management models} $\{h_k:k=1,2, \ldots,K\}$ of size $K \in \mathbb{N}$ is a collection of models that returns a policy by combining the policy derived from each model.
\label{def:3}
\end{definition}

By randomly selecting a part of a larger data set, cash managers can use alternative cash flow data sets to build a collection of models.

\begin{definition}
A \textbf{random cash management set} (RCMS) is an ensemble of models where each model $h_k(b_0,\boldsymbol{f}_k,\alpha)$ is trained by randomly selecting a subsequence $\boldsymbol{f}_k$ belonging to a larger cash flow sequence $\boldsymbol{f}$.
\label{def:4}
\end{definition}

Given an initial condition $b_0$, a context $\alpha$, and a cash flow data set $\boldsymbol{f}$, we next present method to fit a model of the Miller and Orr type as an instance of a general model $h(\boldsymbol{f},b_0,\alpha)$.

\subsection{Fitting bound-based cash management models\label{sec:fitting}}

\begin{figure}[!htb]
\centering
\includegraphics[width=1\textwidth]{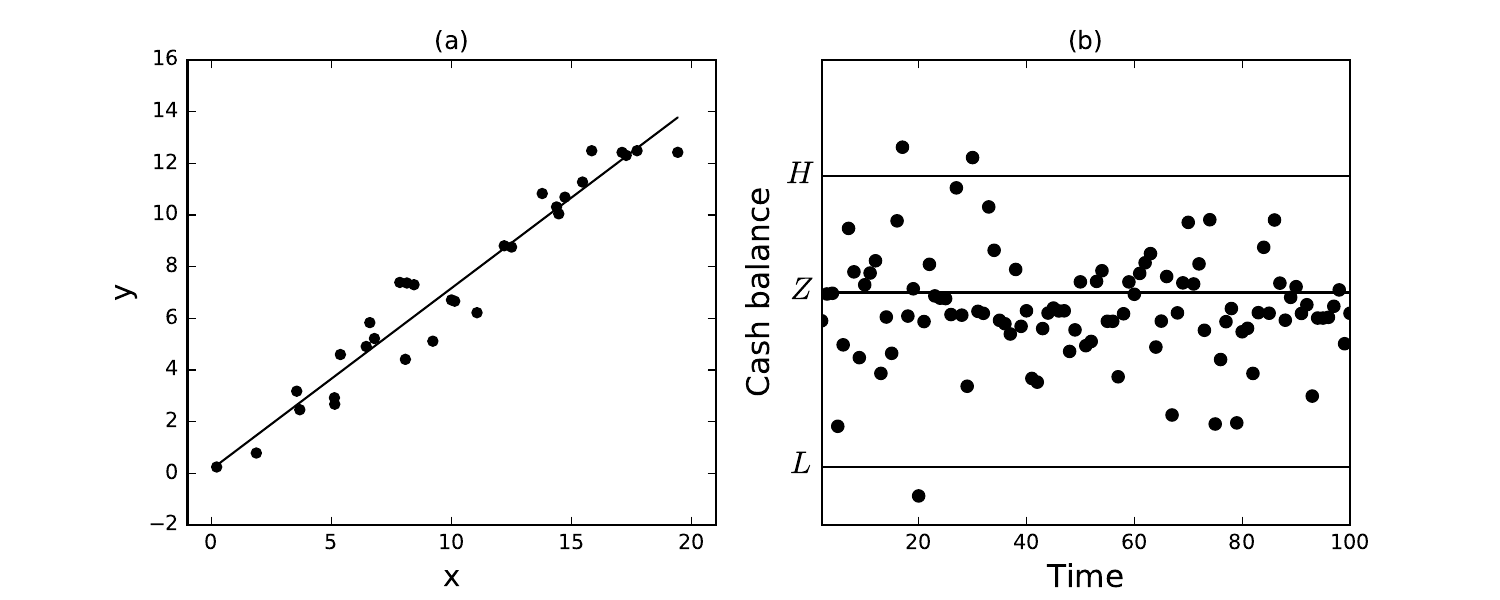}
\caption{\label{fig:ols}(a) Fitting a linear model to data; (b) Fitting a bound-based model to data.}
\end{figure}

Recall from the introduction that we here follow a data-driven approach similar to those used in machine learning to fit models to data. The underlying idea behind the proposed methodology is in line with the main stream of statistical learning approaches ranging from regression analysis to non-parametric estimation. Similarly to linear regression depicted in Figure \ref{fig:ols}~(a), we here use mixed integer linear programming to produce the optimal bound-based model that minimizes the sum of holding and transactions costs introduced in Section \ref{sec:costs} for a given data set of cash flows. The most remarkable difference is that instead of obtaining a set of regression coefficients, we obtain a set of control bounds ready to be used by cash managers to deploy a control policy as shown in Figure~\ref{fig:ols} (b).

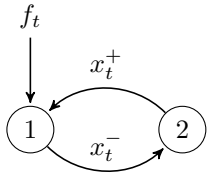
\begin{figure}[htb]
\centering
\begin{tikzpicture}[bend angle=45,node distance = 1.5cm]
\node[place] (account)  {$1$};
\node [text centered, above of=account] (f1) {$f_{t}$}
	edge[post] (account);  
\node[place] (invest_account) [right of= account]  {$2$}
	edge[post, bend right] node[above] {$x_t^+$} (account)
    edge[pre, bend left] node[above] {$x_t^-$} (account);
\end{tikzpicture}
\caption{\label{fig:introex}The common two-assets setting in the cash management problem.}
\end{figure}

In order fit a model $h(\boldsymbol{f},b_0,\alpha)$, we next reformulate objective function \eqref{eq:obj3} with the sum of holding and transaction costs as a mixed integer linear function. To this end, we transform the common two-assets setting shown in Figure \ref{fig:system} into an equivalent configuration as depicted in Figure~\ref{fig:introex}. Let $x_t$ be the difference between control actions $x_t = x_t^+ - x_t^-$ at account~1, with $x_t^+$ and $x_t^-$ being non-negative real numbers. In this setting, the transfer cost function in equation~\eqref{eq:TransferFunction} can be expressed as follows:
\begin{equation}
\Gamma(x_t) = \gamma_0^+ \cdot  z_t^+   + \gamma_1^+ \cdot  x_t^+ +\gamma_0^-  \cdot  z_t^-  + \gamma_1^- \cdot x_t^- 
\label{eq:lineartranscost}
\end{equation}
where $z_t^+, z_t^- \in \{0,1\}$ are binary auxiliary variables satisfying:
\begin{equation}
z_t^+ + z_t^- \leq 1
\end{equation}
\begin{equation}
x_t^+ \leq M \cdot z_t^+
\label{eq:zconstpos}
\end{equation}
\begin{equation}
x_t^- \leq M \cdot z_t^-
\label{eq:zconstneg}
\end{equation}
where $M$ is a very large number. A similar approach can be followed to linearize the holding/penalty cost function in equation \eqref{eq:HoldingShortageCost}. For simplicity, we assume $u=\infty$ and $\Lambda(b_t)=v \cdot b_t$ to restrict ourselves to the usual situation in which cash managers discard policies with negative balances due to high penalty costs. Then, we can rewrite the cost function in equation \eqref{eq:DailyCostFunction} as follows:
\begin{equation}
c(x_t) = \gamma_0^+ \cdot  z_t^+   + \gamma_1^+ \cdot  x_t^+ +\gamma_0^-  \cdot  z_t^-  + \gamma_1^- \cdot x_t^- + v \cdot b_t.
\end{equation}

Furthermore, we must also rewrite the law of motion in equation \eqref{eq:cont} as:
\begin{equation}
b_t = b_{t-1} + f_t + x_t^+ - x_t^-. 
\label{eq:cont2}
\end{equation}

According to the Miller and Orr policy described in equation \eqref{eq:TransferMiller}, positive transactions $x_t^+$ occur when bound $L$ is reached. Thus, $z_t^+=1$ when $b_{t-1}+f_t \leq L$, and the amount transferred is given by $x_t^+ = Z - b_{t-1} - f_t$. This can be expressed by the following linear constraints:
\begin{equation}
b_{t-1} + f_t - L \leq M (1 - z_t^+)
\end{equation}
\begin{equation}
-M (1 - z_t^+) \leq x_t^+ - Z + b_{t-1} + f_t\leq M (1 - z_t^+).
\end{equation}

Furthermore, negative transactions $x_t^-$ occur when bound $H$ is reached. Thus, $z_t^-=1$ when $b_{t-1} + f_t \geq H$, and the amount transferred is given by $x_t^- = b_{t-1} + f_t - Z$. This can be expressed by the following linear constraints:
\begin{equation}
H - b_{t-1} - f_t \leq M (1 - z_t^-)
\end{equation}
\begin{equation}
-M (1 - z_t^-) \leq x_t^- + Z - b_{t-1} - f_t \leq M (1 - z_t^-).
\end{equation}

A third group of conditions must hold when the cash balance is between bounds $L$ and $H$. Thus, when $z_t^+=0$ and $z_t^-=0$, no transaction occurs. This can be expressed by the following linear constraints:
\begin{equation}
L - b_{t-1} - f_t \leq M (z_t^+ + z_t^-)
\end{equation}
\begin{equation}
b_{t-1} + f_t - H \leq M (z_t^+ + z_t^-)
\end{equation}
\begin{equation}
-M \cdot z_t^+  \leq x_t^+ \leq M \cdot z_t^+ 
\end{equation}
\begin{equation}
-M \cdot z_t^- \leq x_t^- \leq M \cdot z_t^-. 
\end{equation}

As a result, given an initial cash balance $b_0$ and a sequence of cash flow observations $\{f_t:t=1,2,\ldots, N\}$ as a given data set, we are in a position to elicit the set of optimal control bounds for policies of the Miller and Orr type by solving the following mixed integer linear program:
\begin{equation}
\operatorname{min} \hspace{2mm} \sum_{t=1}^N \left[ \gamma_0^+ \cdot z_t^+  + \gamma_1^+ \cdot  x_t^+ +  \gamma_0^-  \cdot z_t^-  + \gamma_1^- \cdot x_t^- + v \cdot b_t \right]
\label{eq:lp_miller}
\end{equation}
subject to:
\begin{equation}
b_t = b_{t-1} + f_t + x_t^+ - x_t^-   
\label{eq:miller_cont}
\end{equation}
\begin{equation}
b_{t-1} + f_t - L \leq M (1 - z_t^+)
\label{eq:bound_cont}
\end{equation}
\begin{equation}
-M (1 - z_t^+) \leq x_t^+ - Z + b_{t-1} + f_t \leq M (1 - z_t^+)
\end{equation}
\begin{equation}
H - b_{t-1} - f_t \leq M (1 - z_t^-)
\end{equation}
\begin{equation}
-M (1 - z_t^-) \leq x_t^- + Z - b_{t-1} -f_t \leq M (1 - z_t^-)
\end{equation}
\begin{equation}
L - b_{t-1} - f_t \leq M (z_t^+ + z_t^-)
\end{equation}
\begin{equation}
b_{t-1} + f_t - H \leq M (z_t^+ + z_t^-)
\end{equation}
\begin{equation}
-M \cdot z_t^+  \leq x_t^+ \leq M \cdot z_t^+ 
\end{equation}
\begin{equation}
-M \cdot z_t^- \leq x_t^- \leq M \cdot z_t^- 
\end{equation}
\begin{equation}
z_t^+ + z_t^- \leq 1
\end{equation}
\begin{equation}
x_t^+ \leq M \cdot z_t^+ 
\end{equation}
\begin{equation}
x_t^- \leq M \cdot z_t^- 
\end{equation}
\begin{equation}
b_t \geq b_{min}
\label{eq:bmin}
\end{equation}
where the final decision variables are bounds $L$, $Z$ and $H$ that determine the optimal control policy. At each time step, four additional auxiliary decision variables, two real ($x_t^+$ and $x_t^-$) and two binary ($z_t^+$ and $z_t^-)$ are necessary to solve the problem. Following the recommendations in \cite{ross2002fundamentals} about the Miller and Orr model, we also set a minimum cash balance for precautionary purposes. In practice, setting this value $b_{min}$ is equivalent to set a lower limit for bound $L$.

Note that the problem encoded from equation \eqref{eq:lp_miller} to \eqref{eq:bmin} is a special case of the multistage stochastic problem described from \eqref{eq:sp2} to \eqref{eq:sp2_end} adapted to the characteristics of the \cite{miller1966model} model where random variable $\xi_i = f_t$. On the one hand, objective function \eqref{eq:lp_miller} is the expected cost over some time interval which is equivalent to add the cost of first decision $c_1(x_1)$ to the expected cost of the rest of decisions summarized in $\operatorname{E}_{\xi^2} \left[ Q(\boldsymbol{x},\xi^{\tau}) \right]$ from equation \eqref{eq:sp2}. On the other hand, balance transition equations such as \eqref{eq:miller_cont} or constraint \eqref{eq:bound_cont} relate current state decision variables with previous ones as in constraint \eqref{eq:c3sp2}. As an illustrative example, if vectors $\boldsymbol{x}_i = \left[b_t, x_t^+, x_t^-, z_t^+, y_t\right]^\prime$ and $\boldsymbol{x}_{i-1} = \left[b_{t-1}, x_{t-1}^+, x_{t-1}^-, z_{t-1}^+, y_{t-1}\right]^\prime$ summarize two subsets of decision variables at two consecutive time steps as in equation \eqref{eq:c3sp2}, where $y_t$ and $y_{t-1}$ are additional non-negative auxiliary variables used to transform an inequality into an equality, we can set $W_i$, $T_{i-1}$ and $\boldsymbol{h}_i$ as follows:
\begin{equation}
W_i = \left[ \begin{array}{rrrrr}
1 & -1 &  1 & 0 & 0\\ 
0 & 0 & 0 & M & 1 \\
0 & 0 & 0 & 0 & 0 \\
0 & 0 & 0 & 0 & 0 \\
0 & 0 & 0 & 0 & 0 \\
\end{array} \right];
T_{i-1} = \left[ \begin{array}{rrrrr}
-1 & 0 & 0 & 0 & 0 \\ 
1 & 0 & 0 & 0 & 0  \\ 
0 & 0 & 0 & 0 & 0  \\ 
0 & 0 & 0 & 0 & 0  \\ 
0 & 0 & 0 & 0 & 0  \\ 
\end{array} \right];
\boldsymbol{h}_i = \left[ \begin{array}{c}
f_t \\
M -f_t + L\\
0 \\
0 \\
0 \\
\end{array} \right]
\end{equation}
to obtain equation \eqref{eq:miller_cont} by computing:
\begin{equation}
W_i \boldsymbol{x}_i + T_{i-1} \boldsymbol{x}_{i-1} = \boldsymbol{h}_i.
\end{equation}

By considering the complete vector of decision variables, we can follow a similar reasoning to represent the rest of constraints of the formulation encoded from equations \eqref{eq:lp_miller} to \eqref{eq:bmin} as a general stochastic program such as the one described in Section \ref{sec:stochastic} through the use of larger matrices $W_i$, $T_{i-1}$ and vectors $\boldsymbol{h}_i$, which we here omit for economy of space.

The model proposed in this section is based on an observed cash flow data set as a possible realization of an underlying cash flow process. Then, the solution to the program is the best Miller and Orr model in terms of cost adjusted to the observed cash flow. Intuitively, the goodness of the model depends on how representative of the underlying cash flow process is the data used to fit the model. The higher the number of observations $N$ included in the data set used to fit the model, the higher the probability that the model better captures the real characteristics of the underlying cash flow process. For instance, if the sample volatility is a good estimation of the real volatility, the model will handle well cash flows derived from the real process. The rationale is the same that it is behind modern machine learning techniques aiming to obtain models that generalize well the underlying process. However, a balance between representativeness of the model and computational efficiency must be considered.

However, it is important to highlight that the size $N$ of the available data set may limit the utility of this procedure for computational reasons. Due to the presence of binary variables in the minimization problem encoded from equation \eqref{eq:lp_miller} to \eqref{eq:bmin}, computation times may result prohibitive for large data sets. In order to mitigate this effect, we propose ensemble methods and random subsequences to fit cash management models as formally introduced in Definitions \ref{def:3} and \ref{def:4}. Selecting random subsequences of the input data space has been fruitfully used to construct decision tree models \citep{ho1998random,breiman2001random}. Furthermore, ensemble methods are learning algorithms that construct a set of models and then predict by taking a possibly weighted average of their predictions \citep{dietterich2000ensemble}. Note that the random selection of a subsequence from a time indexed cash flow may be performed by different methods. Taking a uniform sample with replacement as in bagging \citep{breiman1996bagging}, or weighting the observations to produce a biased selection similarly to boosting \citep{freund1996experiments}, are suitable procedures.

\subsection{Generalization power of cash management models\label{sec:adapting}}

Another key feature in machine learning is the appropriateness of a particular predictive model to a given data set. The goodness of fit is usually evaluated by the predictive accuracy that refers to how well the model is able to reproduce the data used to fit the model \citep{makridakis2008forecasting}. In this paper, we follow the approach of measuring how well a particular cash management model fits to a given data set by computing the sum of holding and transaction costs. Furthermore, we propose to measure the utility of any model derived from the optimization problem described in Section \ref{sec:fitting} in comparison to a benchmark model. An interesting benchmark model is the trivial strategy of taking no control action. By comparing performances of models to trivial strategies, we are implicitly checking if cash management models are worthwhile along the lines of \cite{daellenbach1974cash}. We can also use alternative cash management models as a benchmark. In the following case study, we use the equations proposed by the Miller and Orr model described in Section \ref{sec:background} for benchmarking purposes. 

As a result, given a cash flow data set $\boldsymbol{f}$, we define the goodness of fit $G(\boldsymbol{f})$ as:
\begin{equation}
G(\boldsymbol{f}) = \frac{C(\boldsymbol{f})}{C_0(\boldsymbol{f})}
\label{eq:genpower1}
\end{equation}
where $C(\boldsymbol{f})$ is the total cost of the fitted model over data set $\boldsymbol{f}$, and $C_0(\boldsymbol{f})$ is the total cost of the benchmark model over the same data set. The lower the value of $G$, the better the model with respect to the benchmark. However, we are usually more interested in the generalization power of models when dealing with data not used to fit the model. In machine learning, the generalization power of predictive models is estimated by the accuracy of the model in terms of the deviation of forecasts from actual values in a test data set not used to fit the model \citep{makridakis2008forecasting,provost2013data}. To this end, the existing data set is usually split in a training set to fit the model (in-sample data) and a test set to evaluate the model (out-of-sample data). Next, we estimate the generalization power of a cash management model by computing its relative performance with respect to a benchmark model over a test set:
\begin{equation}
G(\boldsymbol{f}_{test}) = \frac{C(\boldsymbol{f}_{test})}{C_0(\boldsymbol{f}_{test})}
\label{eq:genpower2}
\end{equation}
where $\boldsymbol{f}_{test}$ is a cash flow data set not used to fit the model. 

A more sophisticated technique for estimating the generalization power is cross-validation. Unlike training and test set splitting, cross-validation estimates the generalization power of a model by performing multiple splits \citep{hastie2009elements,provost2013data}. This method 
estimates the generalization power of the model across the available data set obtaining useful statistics such as the mean and variance of the expected performance.

\section{Case study\label{sec:case}}

In this section, we describe a case study based on a real cash flow data set from an industrial company in Spain that has been recently used in \cite{salas2017empowering}. This data set contains 2717 daily net cash flows on working days covering a period of more than ten years with mean 0.009 and a high variability 0.097 in terms of standard deviation, both figures in millions of euros. In what follows, we first describe a method to estimate the generalization power of the RCMS defined in Section~\ref{sec:definitions}. Next, we study the impact of alternative economic contexts on generalization power, and finally, we explore the influence of the number of cash flow observations to fit a RCMS. All the experiments in this case study are performed on Jupyter Notebooks executed on a CPU Intel Core Duo E8400 at 3~GHz with 4~GB of RAM under operating system Windows 10 Professional 64 bits. Mathematical programs are solved through the Python interface of Gurobi optimization software \citep{gurobi}.

\subsection{Estimating the generalization power of random cash management models\label{sec:estimating}}

Using the common 80/20 \% split to produce a training and a test set would force us to solve a mixed integer linear program with almost 8700 decisions variables. One of the main purposes of proposing ensembles of models such as the RCMS introduced in Section \ref{sec:fitting} is to obtain a ready-to-use model without solving the extensive problem encoded from equation \eqref{eq:lp_miller} to \eqref{eq:bmin} for a large sequence of observations $f_1, f_2, \ldots, f_N$ when $N$ is large as it is the case of our data set with $N=2717$ cash flows. In order to support this approach with computational results, we next evaluate efficiency in terms of the required computing time to solve the problem for a range of different size samples in which $n \ll N$ holds. The results obtained discouraged us from testing larger values of $n$ since we think that longer computing times are not acceptable in practice from a cash management point of view.

\begin{figure}[!htb]
\centering
\includegraphics[width=0.75\textwidth]{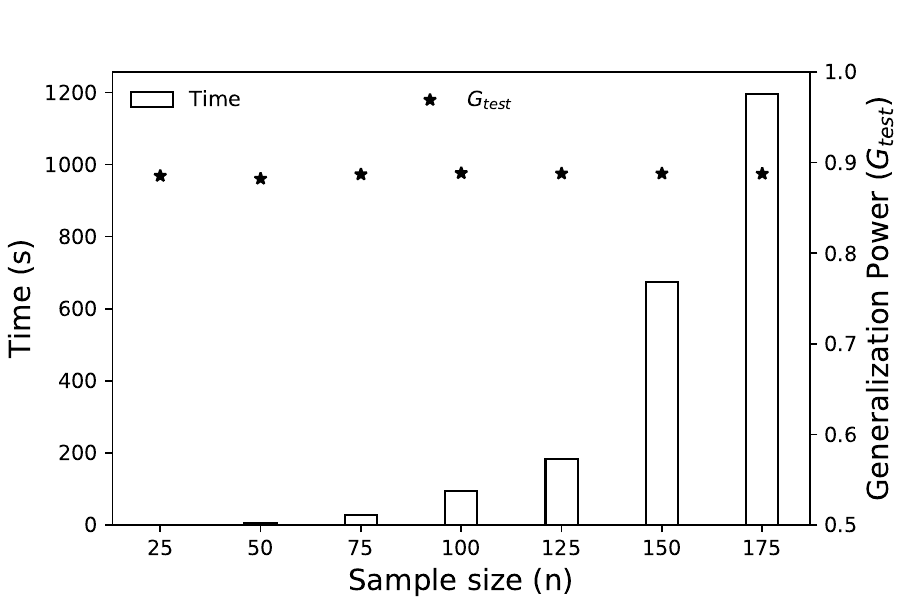}
\caption{\label{fig:time}Run time for different size samples.}
\end{figure}

In Figure \ref{fig:time}, we show the average run time required to solve real instances of problem \eqref{eq:lp_miller}-\eqref{eq:bmin} using a state-of-the-art integer optimization solver such as Gurobi for different size samples. We also depict generalization power $G(\boldsymbol{f}_{test})$ computed using expression \eqref{eq:genpower2} to evaluate the trade-off between efficiency and generalization power of the RCMS. The results in Figure \ref{fig:time} show that run times rapidly increase with the number of observations used to fit the model. However, the generalization power of models remains remarkably stable. As a result, we can reasonably infer that solving the extensive formulation for $N=2717$ observations would lead to prohibitive computing times. To overcome this drawback, we follow the strategy of using a relatively low number of observations fit a good bound-based model. Thus, we here recommend the use of RCMS as a suitable method to find a balance between efficiency and generalization power of bound-based models. In what follows, we use RCMS to fit a cash management model of the Miller and Orr type to the data provided by the company. To this end, we proceed as detailed in Algorithm \ref{GeneralAlg}. 
\\

\begin{algorithm}[H]
 \textbf{Input:} Cash flow data set $\boldsymbol{f}$; initial balance $b_0$; number $n$ of observations to fit the model; number of randomly trained models $K$; cost structure $\alpha$; benchmark model $h_0$; and training/test set rate $r$\;  
 \textbf{Output:} Generalization power for a cash management model\;
 Split $\boldsymbol{f}$ in $\boldsymbol{f}_{train}$ and $\boldsymbol{f}_{test}$ according to $r$\;
 \For{$k=1,2,\ldots,K$}
  {
  Randomly sample $n$ observations from $\boldsymbol{f}_{train}$ to obtain $\boldsymbol{f}_k$\;
  Estimate the model $h_k(b_0,\boldsymbol{f}_k,\alpha)$ by solving program \eqref{eq:lp_miller}-\eqref{eq:bmin} with $N=n$\;
  }
  Obtain model $h(b_0,\boldsymbol{f}_{train},\alpha)$\ by averaging parameters of $h_k$\;
  Compute $G(\boldsymbol{f}_{test})=C(\boldsymbol{f}_{test})/C_0(\boldsymbol{f}_{test})$ from the cost of deploying $h$ and $h_0$\;
 \caption{Generalization power estimation for cash management models}
 \label{GeneralAlg}
\end{algorithm}
\vspace{5mm}

Note that Algorithm \ref{GeneralAlg} can be used for a single generalization power estimate, but it can be replicated as many times as needed for cross-validation. In this case study, we use a training set wit the first 80\% of the observations and a test set with the remaining 20\% since we aim to test the utility of the model with the most recent data. As an illustrative example, let us consider the following cost structure selected from those proposed in \cite{da2014evolutionary}: 
\begin{equation}
\alpha_1 = \left(\gamma_0^+=20 \text{\euro},\gamma_0^-=20 \text{\euro},\gamma_1^+=0.01\%,\gamma_1^-=0.01\%,v=0.02\%,u=\infty\right).
\label{eq:alpha_1}
\end{equation}

In this case study, we use as a benchmark the Miller and Orr model derived from the application of equations \eqref{eq:Z} and \eqref{eq:H} from Section \ref{sec:background}. According to the recommendations in \cite{ross2002fundamentals}, we set a lower bound $L$ for precautionary purposes. A suitable way to do it is setting a value proportional to the empirical standard deviation of cash flows along the lines of \cite{ben1999robust,ben2009robust} for robust optimization. Then, we set:
\begin{equation}
L = \delta \cdot \sigma
\label{eq:lowerbound}
\end{equation}
where $\sigma$ is the empirical standard deviation of the cash flow data set and $\delta$ is a parameter reflecting the attitude towards risk of cash managers so that the higher the value of $\delta$, the more averse to risk they are. Let us consider as abnormal cash flows those with absolute value above five standard deviations as recommended in \cite{gormley2007utility}. Then, we set $\delta=5$ for precautionary purposes in order to ensure that only abnormal cash flows (only 0.25\% of the observations in this data set) may result in a negative cash balance.

In a first numerical example, we aim to compare the generalization power achieved by our RCMS from Section \ref{sec:definitions} with respect to the Miller and Orr equations described in Section \ref{sec:background}. To this end, we first split the whole cash flow data set in a training set with the first 80\% of the observations and a test set with the remaining 20\%. From the standard deviation of cash flows in the training set and cost structure $\alpha_1$ in tuple \eqref{eq:alpha_1}, we obtain a Miller and Orr model using equations \eqref{eq:lowerbound}, \eqref{eq:Z} and \eqref{eq:H} to obtain $(L_0,Z_0,H_0) = (0.48, 0.57, 0.75)$, figures in millions of Euros. Starting at an initial stable cash balance for both models equal to $Z$ from equation \eqref{eq:Z}, we use Algorithm \ref{GeneralAlg} with $K=20$ and $n=25$ to obtain a RCMS with $(L,Z,H) = (0.46, 0.50, 0.64)$ and $G(\boldsymbol{f}_{test})=0.88$. Since the generalization power is below one, our RCMS performs better than the benchmark. Note also that only 25 samples from the training set and 20 randomly trained models are enough for our RCMS to reduce the cost with respect to the Miller and Orr equations evaluated over the test set.

\subsection{Generalization versus economic context}

Cash managers may be interested in analyzing the impact of alternative economic contexts on the generalization power of RCMS. In this section, we compare the generalization power of a RCMS (with $K=20$ and $n=25$) to the Miller and Orr benchmark by applying Algorithm \ref{GeneralAlg} to different cost structures. In addition to cost structure $\alpha_1$ described in tuple \eqref{eq:alpha_1}, we present in Table \ref{tab:contexts} the results obtained for different combinations of holding and transaction costs. More precisely, we consider the case when: fixed transaction costs are doubled ($\alpha_2$); variable transaction costs are set to zero ($\alpha_3$); fixed transaction costs are doubled and variable transaction costs are set to zero ($\alpha_4$); holding costs are doubled ($\alpha_5$); both fixed transaction and holding costs are doubled and variable transaction costs are set to zero ($\alpha_6$); variable  transaction costs are doubled ($\alpha_7$); both fixed and variable transaction costs and also holding costs are doubled ($\alpha_8$). As in the example in Section \ref{sec:estimating}, we assume $u =\infty$, for all contexts in Table \ref{tab:contexts}.

\begin{table}[!htb]
  \centering
  \caption{Generalization power of a RCMS with $K=20$ and $n=25$ for different economic contexts}
    \begin{tabular}{lrrrrrr}
    \hline
    Context & $\gamma_0^+$(\euro) & $\gamma_0^- $(\euro) & $\gamma_1^+(\%)$ & $\gamma_1^-(\%)$ & $v(\%)$ & $G(\boldsymbol{f}_{test})$ \\
    \hline
    $\alpha_1$ & 20    & 20    & 0.01  & 0.01  & 0.02 & 0.88 \\
    $\alpha_2$ & 40    & 40    & 0.01  & 0.01  & 0.02 & 0,89 \\
    $\alpha_3$ & 20    & 20    & 0     & 0     & 0.02 & 0,85 \\
    $\alpha_4$ & 40    & 40    & 0     & 0     & 0.02 & 0,85 \\
    $\alpha_5$ & 20    & 20    & 0.01  & 0.01  & 0.04 & 0,88 \\
    $\alpha_6$ & 40    & 40    & 0     & 0     & 0.04 & 0,86 \\
    $\alpha_7$ & 20    & 20    & 0.02  & 0.02  & 0.02 & 0,89 \\
    $\alpha_8$ & 40    & 40    & 0.02  & 0.02  & 0.04 & 0,88 \\
    \hline
    \end{tabular}%
  \label{tab:contexts}%
\end{table}%

According to equations \eqref{eq:genpower1} and  \eqref{eq:genpower2}, the lower the value of $G(\boldsymbol{f}_{test})$, the better the performance of the RCMS with respect to the Miller and Orr benchmark. Since the generalization power is quite stable, we can conclude that the impact of changes in the economic context is reduced.

\subsection{Generalization versus number of observations used to fit a RCMS}

An interesting additional exercise consists in determining the minimum number of samples that allow our RCMS to improve the generalization power of the benchmark. We may be also interested in the number of samples from which our RCMS does not produce any further improvement or even worsen its performance. Note that in the limit, when the size of the sample approaches the size of the training set, the performance of our RCMS and the Miller and Orr benchmark should be very similar. We answer these questions by means of a learning plot mapping the number of samples $n$ used to train our RCMS to the generalization power $G$ over the test set.

\begin{figure}[!htb]
\centering
\includegraphics[width=0.75\textwidth]{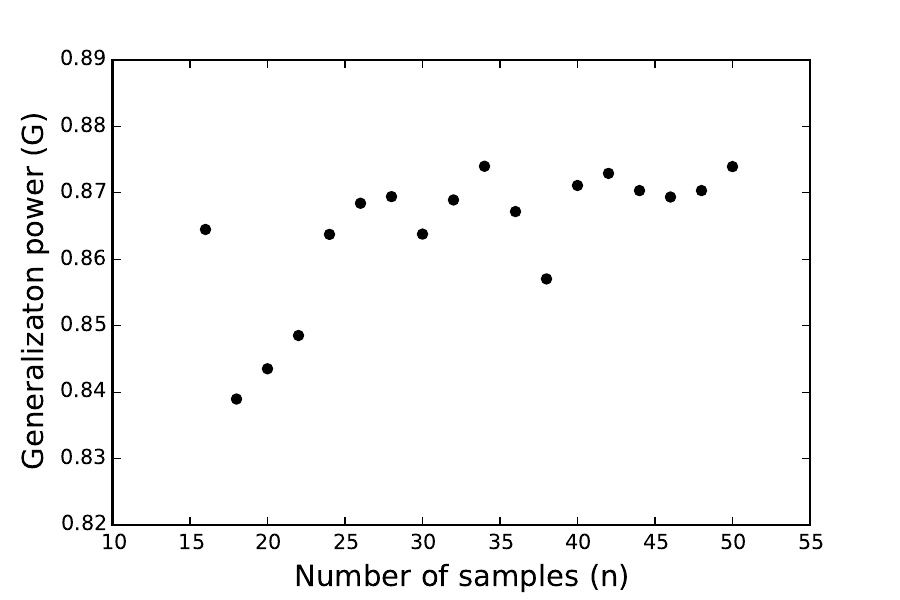}
\caption{\label{fig:learning} Learning plot for random cash management models.}
\end{figure}

In Figure \ref{fig:learning}, we represent the generalization power for ensembles of 20 randomly trained cash management models for different small sample sizes of the training set in steps of two. Sizes below 16 produce results much worse than the benchmark that we remove from the plot for scale reasons. From the analysis of this plot, we can highlight three interesting points: (i) our RCMS consistently outperforms the Miller and Orr benchmark for a wide range of small samples; (ii) the best sample size seems to be around sizes from 18 to 22 observations; and (iii) there is a slight increasing trend in the generalization power with the sample size.

\section{Concluding remarks\label{sec:conclusions}}

Organizations can leverage optimization models to improve data-driven decision-making in finance. Most cash management models are based on the assumption of a particular underlying cash flow process usually assumed to be independent, stationary and Gaussian. In this paper, we relax this strong assumption to fit cash management models to data by relying on both recent machine learning techniques and mathematical programming.

In an attempt to extract useful knowledge from data, we describe a general procedure based on mixed integer linear programs derived from a general stochastic programming approach to elicit the best bound-based model from a given cash flow data set. Our approach is suitable for a wide range of organizations since we use net cash flows that summarize an arbitrary number of flows in a single figure per time step as an input to the model. To reduce the computational effort required to solve large mixed integer linear programs, we also introduce the concept of ensembles of random cash management models. These ensembles are built by randomly selecting a subsequence of the original cash flow data set. Interestingly, the results from a case study using real data show that a small sample size is enough to fit better bound-based models than a benchmark based on the whole training set. These results must encourage cash managers to find a balance between computational effort and generalization power of cash management models deployed to deal with real cash flows. 

It is also important to highlight that our approach can be applied to a variety of data sets including past cash flow observations, draws from a particular probability distribution or even forecasts. Summarizing, we show that stochastic and mixed integer linear programming can be used to fit cash management models to data as way to solve the cash management problem without making any assumption on the available data. Natural extensions of our work may explore more sophisticated methods to randomly train cash management models.

\bibliography{biblio}

\end{document}